\documentstyle[12pt] {article}
\title{\bf Stochastic loop equations}
\author {D.V.ANTONOV \thanks{E-mail:
antonov@pha2.physik.hu-berlin.de, antonov@vxitep.itep.ru.}
\\
{\it Institute of
Theoretical and Experimental Physics,}\\
{\it B.Cheremushkinskaya, 25, 117 218, Moscow, Russia}\\
{\it and}\\
{\it Institut f\"ur Elementarteilchenphysik, Humboldt-Universit\"at,}\\
{\it Invalidenstrasse 110, D-10115, Berlin, Germany}}
\date{}
\begin{document}
\maketitle
\vspace{1cm}

\newcommand{\be} {\begin{equation}}
\newcommand{\ee} {\end{equation}}

\vspace{1cm}
\centerline{\bf {Abstract}}

\vspace{3mm}
Stochastic quantization is applied to derivation of the equations for
the Wilson loops and generating
functionals of the Wilson loops
in the $N=\infty$ limit. These equations are treated both in
the coordinate and momentum representations. In the first case
the connection of the suggested approach with the problem of
random closed contours and supersymmetric quantum mechanics is established,
and the equation for the Quenched Master Field Wilson loop
is derived. The regularized version of one of the obtained equations
is presented and applied to derivation of the equation for the
bilocal field correlator. The momentum loop dynamics is also
investigated.

\newpage
{\large \bf 1. Introduction}

\vspace{3mm}
One of the possible ways of investigation of QCD is its reformulation
in the space of all possible contours (the so-called loop space), when
one considers gauge fields as chiral fields, defined on this space$^{1,2}$.
Loop equations (for a review see$^{3}$) yield natural description of the
confining phase of a gauge theory in terms of elementary excitations, which
are nothing, but interacting closed strings$^{4}$, and the main purpose
of these equations is to choose in the large-$N$ limit those of the free
string theories, which describe gauge fields. This problem is not completely
solved up to now, though several attempts to this direction are done (see
for example$^{5}$). Loop equations are known to be nonlinear 
Schwinger equations in functional derivatives, formulated, generally speaking,
for the generating functionals of the Wilson averages$^{6}$, and that is why
the solution of these equations is a difficult mathematical problem, which
nowadays is disentangled only in two dimensions$^{7}$ (notice also, that
recently in 1+1 QCD the average of arbitrary number of Wilson loops on an 
arbitrary two-dimensional manifold was evaluated, and the relation to the
string theory of maps was established$^{8}$).

Recently alternative gauge-invariant equations were derived$^{9}$ using the
stochastic quantization method$^{10}$ (for a review see$^{11}$), and
 investigated
in various field theories$^{12}$. These equations are written not for Wilson
loops, but for field correlators and therefore are closely connected to the
Method of Vacuum Correlators$^{13,14}$ (for a review see$^{15}$). The
mathematical structure of the obtained equations is simpler than the structure
of the loop equations, since the equations for correlators are just 
evolutionary integral-differential Volterra type-II ones. The derived system
of equations for correlators is in some sense similar to the
Bogolubov-Born-Green-Kirkwood-Ivon chain of equations
for the Green functions in
statistical mechanics$^{16}$, connecting correlators with various number
of fields, and therefore the complete information about the field theory 
under consideration is contained only in the full infinite set of equations.
In order to obtain a closed system of equations one needs to exploit some
kind of approximation. The irreducible field correlators (the so-called
cumulants $^{9,12,13,14,15,17}$ vanish, when any distance between two points,
in which the fields in the cumulant are defined, increases and therefore it
occurs natural to cut the obtained infinite chain of equations, using as
a parameter of the approximation the number of fields in the cumulant (see
discussion in$^{9}$). For example, in$^{9,12}$ the so-called bilocal 
approximation$^{13,14,15}$, when all the cumulants higher than quadratic
were neglected (which corresponded to the Gaussian stochastic ensemble of
fields$^{17}$), was exploited in order to investigate the obtained equations
perturbatively both in the standard, non-gauge-invariant, and gauge-invariant
ways, perform stochastic regulari-\\
zation, separate perturbative gluons$^\prime$
contributions, include external matter fields and apply the suggested approach
to quantization of classical solutions.

The aim of this paper is to apply the stochastic quantization method to
derivation of equations for the Wilson loops in the $N=\infty$ limit
alternative to the familiar loop equations. After that, using 
the cumulant expansion$^{9,12,13,14,15,17}$, one may get equations for
correlators.

There are two types of equations for the Wilson loops, we are going
to present.
The equation of the first type, which will be investigated in
Sections 2 and 3,
is the evolutionary equation of the heat transfer form with the functional
Laplacian standing for the ordinary one. This equation is true in the
limit $N=\infty$ for the arbitrary values of the Langevin time. It will
be derived in Section 2 for a single non-averaged Wilson loop and then
generalized to the case of the generating functional for the Wilson loops.
After that we smear the functional Laplacian in the obtained equation, using
the method, suggested in$^{18}$, and reduce the problem to the integral 
Volterra type-II (by the Langevin time) equation. It will be shown that the
action, over which we shall perform the averaging in order to
invert the smeared 
functional Laplacian (which is the action of the Euclidean harmonic oscillator
at finite temperature$^{18}$), after performing the limiting procedure, which
restores the reparametrization invariance, has a Gaussian form, and therefore
for the problem under consideration the Langevin
equation, where the role of the 
Langevin time plays the parameter of a contour, may be written. Thus we come
to the problem of random closed paths$^{2}$, which due to the fact that it
may be described by the Langevin equation with the Langevin time being the
proper time in the ensemble of contours is possible to be reduced$^{11}$ to
the supersymmetric quantum mechanical problem$^{19}$, so that the dynamics of
this system is governed by the two Fokker-Planck Hamiltonians$^{20,21}$, both
of which survive (in contrast to the case considered in$^{20}$), since one 
should not take the limit of the Langevin time tending to infinity. Also the
equation for the Wilson loop of the Quenched Master Field$^{22}$ is derived 
in Section 2, and it is discussed that due to the discontinuities
of the Wilson
loop in the momentum representation$^{23}$ only the quenching prescription
in the Langevin time direction yields nontrivial equation.

In Section 3 we rewrite the equation for the non-averaged Wilson loop
in the momentum representation, using the methods suggested in$^{23,24}$.
We show that the role of the extra proper time introduced in$^{23}$, in which
the Wilson loop propagates and splits, is played by the Langevin time.

In Section 4 we investigate the equation of the second type, which is also
true only in the $N=\infty$ limit, but, in contrast to the equation of the
first type, only in the asymptotical regime, when the Langevin time tends
to infinity. This is the equation 
for the averaged Wilson loops, where the averaging is
performed over the stochastic
Gaussian noise fields, which is known to be equivalent at the Langevin time
tending to infinity to the averaging with the gluodynamics action weight. 
However, it should be mentioned that since we work in the $N=\infty$ limit 
and neglect all the insertions of the stochastic Gaussian noise fields 
into the Wilson loop, this equation does 
not reproduce planar graphs. Up to now we do not know an equation for the 
averaged Wilson loops alternative to the familiar loop one, which reproduce 
correctly the planar graphs in the large-$N$ limit, where the functional 
Laplacian may be replaced by the operator of differentiation by the Langevin
time.

The equation obtained is then also generalized to the case of the generating 
functional$^{6}$ for the averaged Wilson loops, which provides Veneziano
topological expansion$^{25}$, and rewritten in the momentum representation.
Then we regularize the equation in the coordinate representation, using
the method suggested in$^{26}$, after which it occurs to be defined on the
space of smooth closed contours, to which one can apply the nonabelian
Stokes theorem and the cumulant
expansion. While integrated over the Langevin time $t$, this
ultraviolet-regularized equation with the
properly chosen $t$-dependent momentum cut-off yields the expression
for the one-gluon-exchange diagram, where the perturbative
gluon$^\prime$s propagator is written in
the Feynman-Schwinger path integral representation$^{14}$, and the Langevin
time plays the role of the Schwinger proper time. Finally we use the obtained
regularized equation in order to derive the equation for the
bilocal correlator,
whose Kronecker structure is supposed to be given.

The main results of the paper are summarized in the Conclusion.

\vspace{6mm}
{\large \bf 2. Investigation of the heat transfer type equation
in the coordinate representation}

\vspace{3mm}

We shall start with the Langevin equation$^{10,11}$

$$\dot A_\mu^a=(\nabla_\lambda F_{\lambda\mu})^a-ig\eta_\mu^a,\eqno (1)$$
where $\nabla_\mu=\partial_\mu+[A_\mu,\cdot]$ is the adjoint covariant
derivative, $F_{\mu\nu}=\partial_\mu A_\nu-\partial_\nu A_\mu+[A_\mu,
A_\nu]$ is a strength tensor of the gluonic field, $g=\sqrt{\frac{\lambda}
{N}}$, where $\lambda$ is the bare coupling, which remains finite while
performing the $\frac{1}{N}$ expansion, and $\eta_\mu^a$ is a stochastic
Gaussian noise field.

Let us introduce the following loop functional of the Stokes type

$$\Psi_\eta (C,t)=\frac{1}{N}tr T~ P exp\oint\limits_C^{} dz_\mu
\left(A_\mu (z,t)+ig\int\limits_0^t dt^\prime \eta_\mu(z,t^\prime)\right)
\equiv$$

$$\equiv\frac{1}{N} tr \lim_{\Delta t_i, \Delta z_j\to 0}\prod\limits_i^{} 
\left( 1+\Delta t_i \sum\limits_j^{}(\Delta z_j)_\mu (2A_\mu (z_j, t_i) 
\delta(t_i-t)+ig\eta_\mu(z_j, t_i))\right).$$
Here in addition to the usual $P$-ordering we have introduced $T$-ordering 
because of the second term in the exponent, which contains $t$-integration.

Differentiating $\Psi_\eta (C,t)$ by the Langevin time one gets 
by virtue of (1) 
the following equation

$$\dot \Psi_\eta (C,t)=\frac{1}{N}tr \oint\limits_C^{} dx_\mu (\nabla
_\lambda F_{\lambda\mu}(x)) T~P exp \oint\limits_C^{}dz_\mu\left(
A_\mu(z,t)+ig\int\limits_0^t dt^\prime \eta_\mu (z,t^\prime)\right)=$$

$$=\frac{1}{N} tr\oint\limits_C^{} d x_\mu (\nabla_\lambda F_{\lambda
\mu}(x))\Biggl( \phi(C,t)+ig\oint\limits_C^{} dz_\mu \phi(C_{x_0 z
},t)\int\limits_0^t dt^\prime \eta_\mu (z,t^\prime) \phi(C_{z x_0},t)-$$

$$-g^2\oint\limits_C^{} d z_\mu \oint\limits_C^{} d y_\nu \phi(C_{x_0 z},t)
\int\limits_0^t dt^\prime \eta_\mu(z,t^\prime) \phi(C_{zy},t)
\int\limits_0^t dt^{\prime\prime}\eta_\nu(y,t^{\prime\prime})
\phi(C_{y x_0},t)+...\Biggr),\eqno
(2)$$
where $\phi(C_{xy},t)\equiv P~ exp \int\limits_{C_{xy}}^{}
dz_\mu A_\mu(z,t)$, 
and $x_0$ is an arbitrary but fixed point belonging to the contour $C$.

It was proved in$^{6}$ that the $\frac{1}{N}$ expansion corresponded to
the WKB approximation around the ``classical'' field in the effective 
scalar theory in the loop space. We see that this is really the case
according to equation (2) since in the stochastic quantization
method$^{10,11}$ the degree of quantum correction of the grand ensemble
of fields $A_\mu^a$ and $\eta_\mu^a$ is determined through the maximal
number of the noise fields entering the physical quantity under 
consideration, and on the right-hand side of equation (2) any new term
of the $\frac{1}{N}$ expansion gives rise to the additional power of the 
noise field.

In what follows we shall derive and investigate the equation for the
pure ``classical'' field in the loop space
$\Psi(C,t)=\frac{1}{N} tr \phi(C,t)$ corresponding to $N=\infty$. This 
equation reads

$$\dot \Psi(C,t)=\oint\limits_C^{} dx_\mu \partial^x_\lambda
\frac{\delta}{\delta\sigma_{\lambda\mu}(x)} \Psi(C,t),~~
t<+\infty,
\eqno (3)$$
where $\partial^{x(\sigma)}_\lambda\equiv\int\limits_{\sigma-0}^
{\sigma+0} d\sigma^\prime\frac{\delta}{\delta x_\lambda (\sigma^\prime)}$.

Notice that equation (3) is true for an arbitrary
$t<+\infty$, and from the mathematical point of view it is a homogeneous
functional heat transfer equation (which means that the loop space Laplacian
stands there for the ordinary one), while the loop equation in the $N=\infty$
limit is an inhomogeneous functional Laplace one.

Using equation (3) one can easily write an equation for the generating
functional of the Wilson loops

$$Z(J(C))=exp\left(\sum\limits_C^{}J(C)\Psi(C)\right),$$
where the sum over the loops is defined as follows$^{6}$

$$\sum\limits_C^{}f(C)=\int\limits_{+0}^{+\infty}\frac{dT}{T}\int\limits
_{x(0)=x(T)}^{} Dx(\alpha)f(x(\alpha)).$$
This equation reads

$$\frac{\partial}{\partial t} \frac{\delta Z}{\delta J(C)}=
\oint\limits_C^{} dx_\mu \partial^x_\lambda
\frac{\delta}{\delta\sigma_{\lambda\mu}(x)} \frac{\delta Z}
{\delta J(C)}+\sum\limits_{C^\prime}^{}J(C^\prime)\oint\limits_
{C^\prime}^{}dz_\mu \partial^z_\lambda
\frac{\delta}{\delta\sigma_{\lambda\mu}(z)}
\frac{\delta^2 Z}{\delta J(C)\delta J(C^\prime)}\eqno (4)$$
and does not depend on $N$ explicitly. We shall see in Section 4 that 
this dependence appears in the equation for the generating functional
of the averaged Wilson loops, which describes an "open string" with
quarks at the ends. This equation will correspond to the Veneziano
topological expansion$^{6,25}$.

Let us now perform the smearing of the functional Laplacian, standing
on the right-hand side of equation (3), by making use of the method 
suggested in$^{18}$.
It may be done via performing the polygon discretization procedure and
then going to the continuum limit or directly using the continuum smearing
without any reference to the polygon discretization. This smearing implies
that one replaces the functional Laplacian

$$\Delta\equiv\oint\limits_C^{} dx_\mu \partial^x_\lambda
\frac{\delta}{\delta\sigma_{\lambda\mu}(x)}=\int\limits_0^1 d\sigma
\int\limits_{\sigma-0}^{\sigma+0} d\sigma^\prime \frac{\delta}
{\delta x_\mu(\sigma^\prime)} \frac{\delta}{\delta x_\mu(\sigma)}$$
with the smeared one

$$\Delta^{(G)}=\int\limits_0^1 d\sigma~~ v.p.\int\limits_0^1 d\sigma^\prime
G(\sigma-\sigma^\prime)\frac{\delta}{\delta x_\mu(\sigma^\prime)}
\frac{\delta}{\delta x_\mu(\sigma)}+\Delta,\eqno (5)$$
where $G(\sigma-\sigma^\prime)$ is a smearing function, and the first
term on the right-hand side of equation (5), which involves the 
principal-value integral, is an operator of the second order (does not
satisfy the Leibnitz rule in contrast to the operator $\Delta$) and 
reparametrization noninvariant. The
averaging over the loops $\xi(\sigma)$ is defined as follows

$$<F[\xi]>_\xi^{(G)}=\frac{\int\limits_{\xi(0)=\xi(1)}^{} D\xi 
e^{-S}F[\xi]}{\int\limits_{\xi(0)=\xi(1)}^{} D\xi e^{-S}},\eqno (6)$$
where

$$S=\frac{1}{2}\int\limits_0^1 d\sigma\int\limits_0^1 d\sigma^\prime
\left(\xi(\sigma)G^{-1}(\sigma-\sigma^\prime)\xi(\sigma^\prime)\right),
\eqno(7)$$
and $G^{-1}$ is an inverse operator. For the simplest case

$$G(\sigma-\sigma^\prime)=e^{-\frac{|\sigma-\sigma^\prime|}{\varepsilon}},
~~\varepsilon\ll 1$$
the action (7) reduces to the action of the Euclidean harmonic
oscillator at finite temperature$^{18}$

$$S=\frac{1}{4}\int\limits_0^1
d\sigma\left(\varepsilon\dot\xi^2(\sigma)+
\frac{1}{\varepsilon}\xi^2(\sigma)\right),\eqno (8)$$
which may be recovered from the discretized action, when the number
of vertices of the polygon, which approximates the loop, tends to
infinity.
The contribution of the first term on the right-hand side of equation
(5)
is of the order $\varepsilon$ for smooth contours since the region
$|\sigma-\sigma^\prime|\sim\varepsilon$ is essential in the integral
over $d\sigma^\prime$. Hence when $\varepsilon$ tends to zero, this 
term vanishes, reparametrization invariance restores, and $\Delta^
{(G)}\to \Delta$.

Thus equation (3) takes the form

$$\Delta^{(G)}\Psi[x,t]=\dot\Psi[x,t],\eqno(9)$$
where $x=x(\sigma)$ is the position vector of the contour $C$.
Using the equation of motion

$$<\xi_\mu(\sigma)F[\xi]>_\xi^{(G)}=\int\limits_0^1 d\sigma^\prime
G(\sigma-\sigma^\prime)<\frac{\delta F[\xi]}{\delta\xi^\mu(\sigma
^\prime)}>_\xi^{(G)},$$
which follows from the oscillator action (8), one can solve
equation (9)$^{18}$ with the condition $\Psi[0,t]=1$. Integrating over
the Langevin time we arrive at the following equation

$$\int\limits_0^t dt^\prime \Psi[x,t^\prime]+\frac{1}{2}
\int\limits_0^{+\infty}d\gamma \left(<\Psi[x+\sqrt{\gamma}\xi, t]>_\xi^{(G)}-
<\Psi[\sqrt{\gamma}\xi, t]>_\xi^{(G)}\right)=$$
$$=t+\frac{1}{2}\int\limits_0^
{+\infty} d\gamma\left(<\Psi[x+\sqrt{\gamma}\xi, 0]>_\xi^{(G)}
-<\Psi[\sqrt{\gamma}
\xi, 0]>_\xi^{(G)}\right),\eqno (10)$$
which is the integral Volterra type-II one by the Langevin time, and
the
dependence on the initial conditions is put to the right-hand side.

Finally let us go to the limit $\varepsilon\to 0$ in order to restore
the reparametrization invariance of equation (10). Then performing
the rescaling $\zeta=\frac{\xi}{\sqrt{\varepsilon}}$, we see that the 
averaging in both sides of equation (10) becomes Gaussian with the
action $S=\frac{1}{4}\int\limits_0^1 d\sigma \zeta^2(\sigma)$. It
means
that we come to the problem of motion of random closed paths, which
occurs to be described by the Langevin equation, where the role of
the Langevin time plays the parameter of the contour (i.e. the proper time
in the loop space). The problem of random closed contours was investigated 
in$^{2}$, where it was shown that the number of contours of the length
$T$ was given by the formula

$$dN(T)=\frac{dT}{T}T^{-\frac{D}{2}}e^{-\frac{cT}{\epsilon}},$$
where $D$ was the dimension of the space-time, $c$ was some constant,
and $\epsilon$ was a cut-off parameter.

Within our approach we reduced this problem to the Langevin equation,
written in the proper time of the system under consideration.
Therefore
it is equivalent$^{11}$ to the supersymmetric quantum mechanical
problem$^{19}$. Namely if one writes down in components the equation
of the Brownian motion of a classical particle in a heat bath,
which describes the evolution of the position vector of a contour,

$$\frac{\partial x}{\partial\sigma}=-\frac{\delta S}{\delta x}
+\eta,$$
then making the change of variables $\eta\to x$ in the partition 
function for the Langevin dynamics, we obtain the following effective
Lagrangian

$${\cal L}_{eff.}=\frac{1}{2}\dot x^2+\frac{1}{2}\dot x V+\frac{1}{8}
V^2-\bar\psi\left(\frac{\partial}{\partial\sigma}+\frac{1}{2}V^\prime
\right)\psi,$$
where $"~~^{.}~~"\equiv\frac
{\partial}{\partial\sigma}, "~~^\prime~~"\equiv\frac{\delta}{\delta x},
~~V\equiv S^\prime$.
In order to have a closed supersymmetry algebra, one needs to
introduce
an auxiliary field $D$:

$${\cal L}=\frac{1}{2}\dot x^2-\frac{1}{2} D^2-\frac{1}{2} DV-
\bar\psi\left(\frac{\partial}{\partial\sigma}+\frac{1}{2}V^\prime
\right)\psi,\eqno (11)$$
so that the action, corresponding to (11), is invariant under the
supersymmetry transformations

$$\delta x=\bar\varepsilon \psi-\bar\psi \varepsilon,~~
\delta D=\bar\varepsilon \dot \psi+\dot {\bar\psi}\varepsilon,~~
\delta\psi=(\dot x+D)\varepsilon,~~
\delta \bar\psi=\bar\varepsilon(\dot x-D).$$
The dynamics of the system may be shown$^{20,21}$ to be governed by
the two Fokker-Planck Hamiltonians

$$H^\pm=-\frac{1}{2}\frac{\delta^2}{\delta x^2}+\frac{1}{8}V^2
\pm\frac{1}{4}V^\prime,$$
where the Hamiltonian $H^{-}$ corresponds to the propagation forward
in the Langevin time $\sigma$, and $H^{+}$ corresponds to the backward
propagation. The existence of the two Fokker-Planck Hamiltonians is
a consequence of the supersymmetry. However it should be emphasized
that while in the usual causal interpretation of the Langevin
equation, which requires propagation forward in the Langevin time, in
the equilibrium limit only the forward dynamics survives since only
the zero ground state of $H^{-}$ contributes, and $H^{+}$ has strictly
positive eigenvalues$^{20}$, in our model it is not the case, because
one should not take the limit $\sigma\to +\infty$. Therefore the
dynamics of the system is governed by both the forward and backward
Hamiltonians.

To conclude this Section, we obtain the equation for the Quenched
Master Field$^{22}$ Wilson loops. The important property of the Wilson
loops in the momentum representation is the presence of finite
discontinuities induced by the emission and absorption
of gluons$^{23}$

$$\Delta p_\mu (s_i)=p_\mu (s_i+0)-p_\mu(s_i-0),$$
and therefore only quenching in the Langevin time direction is valid

$$A_\mu^{ab}(x,t)=e^{i(p_{5a}-p_{5b})t}\bar A_\mu^{ab}(x),~~
\eta_\mu^{ab}(x,t)=e^{i(p_{5a}-p_{5b})t}\bar \eta_\mu^{ab}(x),$$
where

$$<\bar \eta_\mu^{ab}(x)\bar \eta_\nu^{cd}(y)>=
2\frac{\Lambda_5}{2\pi}\delta^{bc}
\delta^{ad}\delta_{\mu\nu}\delta(x-y).$$
Introducing a new loop functional

$$\Omega (C,t)=\frac{1}{N} tr\oint\limits_C^{} d x_\mu 
P\left(B_\mu(x,t)exp\oint\limits_C^{}d y_\lambda A_\lambda 
(y,t)\right),$$
where
$B_\mu^{ab}(x,t)=i(p_{5a}-p_{5b})
A_\mu^{ab}(x,t),$
we obtain from (3), performing the smearing
of the functional Laplacian as it has been done above, the following
equation for the loop functionals $\Psi$ and $\Omega$:

$$\Psi[x,t]=1-\frac{1}{2}\int\limits_0^{+\infty} d\gamma
\left(<\Omega[x+\sqrt{\gamma}\xi, t]>_\xi^{(G)}-<\Omega[\sqrt{\gamma}
\xi,t]>_\xi
^{(G)}\right).$$

\vspace{6mm}
{\large \bf 3. Heat transfer type equation in the momentum
representation}

\vspace{3mm}
In this Section we shall rewrite equation (3) in the momentum
representation
in order to investigate free momentum loop dynamics in the Langevin
time. During this free motion the loop behaves as a collection 
of free particles, while the loops$^\prime$ collisions, at which
momenta are rearranged between particles, which means that the loops
split, will be considered in the next Section.

The loop functional in the momentum representation is defined as
follows$^{23,24}$:

$$\Psi(P,t)=\int DC exp\left(i\int\limits_C^{}p_\mu dx_\mu\right)
\Psi(C,t).$$
Using the representation

$$\Delta=\int\limits_0^1 ds_1\int\limits_{s_1-0}^{s_1+0} ds_2
\frac{\delta^2}{\delta x_\mu(s_1)\delta x_\mu(s_2)}$$
for the functional Laplacian and following the procedures suggested
in$^{23,24}$, one gets from equation (3) two equations, which describe
the free propagation of the momentum loop:

$$\dot\Psi(P,t)=-\sum\limits_i^{}(\Delta p_\mu(s_i))^2\Psi(P,t),\eqno
(12)$$

$$\dot\Psi(P,t)=-2\int\limits_0^1 ds_1 p_\mu^\prime(s_1)\int\limits
_0^{s_1} ds_2 p_\mu^\prime(s_2) \Psi(P,t).$$
Notice that in equation (12) the reparametrization invariance is
preserved, i.e. when the parameter $s$ changes to $f(s)$, where
$f^\prime(s)>0$, the positions of the discontinuities $s_i$ shift,
but their order and the values of $\Delta p_\mu(s_i)$ remain
the same.

Therefore one may conclude that the role of the extra proper time $H$,
which was introduced in$^{23}$, in which the Wilson loop 
propagates and bits, 
plays the Langevin time $t$. This evolution is however alternative
to the evolution of the loop in the space of random contours,
described
by the supersymmetric quantum mechanics (see the previous Section),
where the role of the proper time played the parameter of the contour.

\vspace{6mm}
{\large \bf 4. An equation for the averaged Wilson loops}

\vspace{3mm}

This Section is devoted to investigation of the equation alternative
to equation (3), which will be written for the averaged Wilson loops.
It can be derived by averaging both sides of equation (3) over the 
stochastic noise fields $\eta_\mu^a$, which in the physical limit
of the Langevin time tending to infinity is equivalent$^{10,11}$ to
the averaging with the gluodynamics action weight. Therefore in the
asymptotical regime $t\to+\infty$ one may use the equation of motion
for the Wilson loop $\Psi(C,t)$, which is nothing, but the loop
equation in the limit $N=\infty$. Thus we obtain the following equation:

$$\dot\Phi(C,t)=\lambda\oint\limits_C^{} dx_\mu~~v.p.\int\limits_C^{}
dy_\mu
\delta(x-y)\Phi(C_{xy},t)\Phi(C_{yx},t),~~t\to+\infty,
\eqno (13)$$
where $\Phi(C,t)=<\Psi(C,t\to+\infty)>_{\eta_\mu^a}$, and the principal value
integral on the right-hand side of equation (13) implies that we
integrate only over those $y_\mu~^\prime$s, which are another points
of the contour $C$ as the point $x$. In other words the right-hand
side of equation (13) does not vanish only when the contour $C$ has
self-intersections.

As it was already mentioned in the Introduction, this equation does 
not reproduce the planar graphs since we have neglected insertions of the 
stochastic noise 
fields into the Wilson loop in the limit $N=\infty$. 
It should be understood as an 
asymptotical at $t$ tending to infinity equation for the ``classical'' 
Wilson average.

Equation (13) may be generalized to the case of the generating
functional ${\cal Z}(C, j)=\frac{1}{N}\frac{\delta~ln<Z>}{\delta
J(C)}$, which is also a functional of the source
$j(C)=\frac{J(C)}{N}$. If $N$ tends to infinity, but the ratio 
$\rho=\frac{N_f}{N}$ is fixed, which corresponds to the Veneziano
topological expansion$^{25}$, generalizing the $\frac{1}{N}$
expansion, then the source $j$ remains finite. The equation 
for the functional ${\cal Z}$ reads

$$\frac{\partial}{\partial t} {\cal Z}(C)=\lambda\oint\limits_C^{}
dx_\mu~~v.p.\int\limits_C^{}dy_\mu\delta(x-y)\left(N{\cal Z}(C_{xy})
{\cal Z}(C_{yx})+\frac{1}{N}\frac{\delta{\cal Z}(C_{xy})}{\delta j
(C_{yx})}\right)+$$
$$+\sum\limits_{C^\prime}^{}j(C^\prime)\oint\limits_{C^\prime}^{}
dz_\mu\partial^z_\lambda\frac{\delta}{\delta\sigma_
{\lambda\mu}(z)}\left(N^2{\cal Z}(C){\cal Z}(C^\prime)+\frac{\delta
{\cal Z}(C)}{\delta j(C^\prime)} \right).$$
This is the Schwinger equation in functional derivatives, which
describes the evolution of the wave functional of the open string
with quarks at the ends (see discussion in$^{6}$).

Let us now rewrite equation (13) in the momentum representation. It
may be done using the procedure, suggested in$^{23}$, and the answer
has the form

$$\dot\Phi(P,t)=-2\lambda\int\limits_0^1 ds_2\frac{\delta}{\delta
p_\mu(s_2)}\int\limits_0^{s_2}ds_1\frac{\delta}{\delta p_\mu(s_1)}
\Phi(P^{(1)},t)\Phi(P^{(2)},t), \eqno (14)$$
where $P^{(1)}$ and $P^{(2)}$ are the parts of $P$ from $s_1$ to $s_2$
and from $s_2$ to $s_1$ respectively:

$$P^{(1)}:~p_\mu^{(1)}(s)=
p_\mu((1-s_{21})s),~0<s<s^\prime,~~p_\mu^{(1)}(s)=
p_\mu((1-s_{21})s+s_{21}),~s^\prime<s<1,
\eqno (15)$$

$$P^{(2)}:~p_\mu^{(2)}(s)=p_\mu(s_1+s_{21}s),~0<s<1, \eqno (16)$$
where $s_{21}=s_2-s_1,~ s^\prime=\frac{s_1}{1-s_{21}}$.

Equation (14) describes the vertex of the splitting of the loop into
two loops $P \to P^{(1)}+P^{(2)}$, so that the momentum $p_\mu (s)$ of
the initial loop is distributed between these two new loops according
to the formulae (15) and (16). As was discussed in$^{23}$, at every
splitting the effective number of degrees of freedom per loop
diminishes,
so that if one approximates the initial loop by a polygon, the number
of vertices of each of the polygons, approximating the loops after the
splitting, will be smaller. At each splitting of the loop the function
$p_\mu(s)$ smothers, and the derivative $p_\mu^\prime(s)$ decreases.
 Finally the loop, propagating in the Langevin time and splitting
during this process, runs out of degrees of freedom and reduces to the
point in the momentum space $p_\mu(s)=p_\mu=const$.

Our next aim is to apply to equation (13) the nonabelian Stokes
theorem and the cumulant
expansion. After that using the bilocal approximation$^{13,14,15}$, we
shall derive the equation for the bilocal correlator, whose Kronecker
structure will be supposed to be given. To this end we perform an
ultraviolet regularization of equation (13), exploiting the method,
suggested in$^{26}$, where this procedure was done with the help of
the heat-kernel regularized Langevin equation. The regularized version
of equation (13) has the form

$$\dot\Phi(C,t)=\lambda\oint\limits_C^{}dx_\mu\oint\limits_C^{}dy_\mu
\int\limits_x^yDrexp\left(-\frac{1}{4}\int\limits_
0^{\Lambda^{-2}}ds \dot
r^2(s)\right)\Phi(C_{xy}r_{yx},t)\Phi(C_{yx}r_{xy},t
), \eqno (17)$$
where $r(0)=x,~r(\Lambda^{-2})=y$, and the momentum cut-off $\Lambda$
may be taken for example to be of
the order of the inverse correlation length of the vacuum$^{15}$
$T_g \simeq 0.2 fm: |\Lambda|\sim\frac{1}{T_g}$. However
 in what follows we
shall use as a momentum cut-off $\frac{1}{\sqrt{t}}$.

It is known$^{27}$ that in the stochastic quantization of gauge
theories it is not necessary to add a gauge-fixing term into the
Langevin equation, since direct iterations in powers of the coupling
constant without introducing ghost fields yield the same results
as the Faddeev-Popov perturbation theory, because the Langevin time $t$
takes the role of a gauge parameter. If one fixes $t$, calculates
gauge-invariant quantities and then goes to the physical limit
$t\to +\infty$, all the lineary divergent in $t$ terms will cancel
each other in the same way as the terms, depending on the gauge
parameter in the framework of the usual approach.

Therefore our choise of the momentum cut-off looks natural from the
point of view of the general principa of the stochastic quantization
method. If we now integrate equation (17) by the Langevin time, 
analitically continuing both sides down from the asymptotics $t\to +
\infty$, then the obtained equation yields the expression for the
second order one-gluon-exchange diagram in the perturbation theory
in the nonperturbative gluodynamics vacuum$^{14}$, where gluonic
propagator is written in the Feynman-Schwinger path integral
representation, and the role of the Schwinger proper time plays the
Langevin time. This perturbative gluon propagates along a regulator
path $r_{xy}$, dividing the surface, swept by the nonperturbative 
gluodynamics string, into two pieces.

Equation (17) is defined on the space of closed smooth contours, and
therefore one can apply to it the nonabelian Stokes theorem and the cumulant
expansion, after which
the integral over the regulator paths, standing on the right-hand side
of this equation, yields the amplitude of propagation of a free gluon.
Let us vary both sides of equation (17), to which we applied the 
cumulant expansion in the bilocal approximation, by the element
of the surface, lieing on the contour $C$, and suppose for simplicity
that the bilocal correlator has the only one Kronecker structure:

$$<F_{\mu\nu}(x,t)\phi(\Gamma_{xy},t) F_{\lambda\rho}(y,t)
\phi(\Gamma_{yx},t)>=
\frac{\hat 1}{N} (\delta_{\mu\lambda}\delta_{\nu\rho}-
\delta_{\mu\rho}\delta_{\nu\lambda})D((x-y)^2,t),$$
where $\Gamma_{xy}$ and $\Gamma_{yx}$ are some contours.
Then we get the following equation for the correlation function
$D$:

$$\int\limits_S^{}d\sigma_{\lambda\rho}(x_2)D((x_1-x_2)^2,t)=
\frac{g^2}{8\pi^2}\frac{\delta}{\delta\sigma_{\lambda\rho}(x_1)}\oint
\limits_C^{}dx_\mu\oint\limits_C^{}dy_\mu\frac{e^{-\frac{(x-y)^2}
{4t}}}{(x-y)^2},$$
where the point $x_1$ belongs to the contour $C$, the surface $S$
is bounded by this contour, and the initial condition $D(x^2,0)=0$
is implied.

\vspace{6mm}
{\large \bf 5. Conclusion}

\vspace{3mm}
In this paper we applied the stochastic quantization method to
derivation of the equations for Wilson loops in the $N=\infty$ limit.
There are two types of equations: first is the heat transfer type
equation (3), which is true for arbitrary values of the Langevin time.
This is the evolutionary equation for the non-averaged Wilson loop. It
was derived in Section 2 and then generalized to the case of the
generating functional for the Wilson loops. After that, applying the
smearing procedure to the functional Laplacian which stands on the
right-hand side of this equation, we reduced the problem to the
Volterra type-II integral equation (by the Langevin time) and
established its connection with the problem of random closed paths.
The latter occured to be described by the supersymmetric quantum
mechanics of a particle in a heat bath, where the role of the proper
time played the parameter of a contour. Therefore the dynamics of the
system of random contours is governed by the forward and backward
Fokker-Planck Hamiltonians.

The quenching prescription may be applied to the obtained equation
only in the Langevin time direction, since in the momentum
representation the Wilson loop possesses discontinuities. The equation
for the Wilson loop of the Quenched Master Field is presented at the
end of Section 2.

In Section 3 the heat transfer type equation was investigated in the
momentum representation, and two alternative equations, describing the
free propagation of the non-averaged Wilson loop in the Langevin time, were
derived. 

In Section 4 an alternative equation for the averaged Wilson loops,
which was also true in the limit $N=\infty$ but 
in the asymptotical regime of the large values of the
Langevin time $t$, was derived and generalized to the case of the
generating functional for the Wilson averages. This equation for the
averaged Wilson loops, while rewritten in the momentum representation,
describes splitting of the loop during its propagation in the Langevin
time. At these splittings the loop loses its degrees of freedom and
finally shrinks to a point in the momentum space.

Then in order to apply to the equation in the coordinate
representation the nonabelian Stokes theorem and the 
cumulant expansion and derive the equation for the
bilocal cumulant, we regularized this equation, using the method
suggested in$^{26}$, after which, choosing for the momentum cut-off
$\frac{1}{\sqrt{t}}$ and integrating over $t$, one may recognize in
the obtained expression the one-gluon-exchange diagram$^{14}$,
contributing to the averaged Wilson loop. Applying to the both sides of
this equation the nonabelian Stokes theorem and the
cumulant expansion and assuming that the bilocal
correlator had a given Kronecker structure, we obtained the equation
for the corresponding coefficient function.

\vspace{6mm}
{\large \bf 6. Acknowledgements}

\vspace{3mm}
I am grateful to Professors Yu.M.Makeenko and Yu.A.Simonov for
useful discussions and to the
theory group of the Quantum Field Theory Department of the Institut f\"ur 
Physik of the Humboldt-Universit\"at of Berlin for kind hospitality. 
The work is supported by Graduiertenkolleg {\it Elementarteilchenphysik}, 
Russian Fundamental Research
Foundation, Grant No.96-02-19184, DFG-RFFI, Grant 436 RUS 113/309/0 
and by the INTAS, Grant No.94-2851.

\newpage
{\large\bf References}

\vspace{3mm}
\noindent
1.~A.M.Polyakov, {\it Phys.Lett.} {\bf B82}, 247 (1979), {\it
Nucl.Phys.} {\bf B164}, 171 (1980).\\
2.~A.M.Polyakov, {\it Gauge Fields and Strings} (Harwood, 1987).\\
3.~A.A.Migdal, {\it Phys.Rep.} {\bf 102}, 199 (1983).\\
4.~K.Wilson, {\it Phys.Rev.} {\bf D10}, 2445 (1974); J.Kogut and 
L.Susskind, {\it Phys.Rev.} {\bf D11}, 395 (1975).\\
5.~A.A.Migdal, {\it Nucl.Phys.} {\bf B189}, 253 (1981).\\
6.~Yu.M.Makeenko and A.A.Migdal, {\it Nucl.Phys.} {\bf B188}, 269 (1981).\\
7.~V.A.Kazakov and I.K.Kostov, {\it Nucl.Phys.} {\bf B176}, 199
(1980); V.A.Kazakov, {\it Nucl.Phys.} {\bf B179}, 283 (1981); 
E.Karjalainen, {\it Nucl.Phys.} {\bf B413}, 84 (1994).\\
8.~D.J.Gross, {\it Nucl.Phys.} {\bf B400}, 161 (1993); D.J.Gross and 
W.Taylor, {\it Nucl.Phys.} {\bf B400}, 181 (1993), {\it Nucl.Phys.} 
{\bf B403}, 395 (1993).\\
9.~D.V.Antonov and Yu.A.Simonov, {\it Int.J.Mod.Phys.} {\bf A11}, 4401 
(1996).\\
10. G.Parisi and Y.-S.Wu, {\it Sci.Sin.} {\bf 24}, 483 (1981).\\
11. P.H.Damgaard and H.H\"uffel, {\it Phys.Rep.} {\bf 152}, 227
(1987).\\
12. D.V.Antonov, {\it JETP Lett.} {\bf 63}, 398 (1996), 
{\it hep-th} /9605044 ({\it Yad.Fiz.}, in
press), {\it hep-th} /9605045 ({\it Yad.Fiz.}, in press).\\
13. H.G.Dosch, {\it Phys.Lett.} {\bf B190}, 177 (1987); Yu.A.Simonov, 
{\it Nucl.Phys.} {\bf B307}, 512 (1988); H.G.Dosch and Yu.A.Simonov, 
{\it Phys.Lett.} {\bf B205}, 339 (1988), {\it Z.Phys.} {\bf C45}, 147 
(1989); Yu.A.Simonov, {\it Nucl.Phys.} {\bf B324}, 67 (1989), {\it
Phys.Lett.} {\bf B226}, 151 (1989), {\it Phys.Lett.} {\bf B228}, 413 
(1989).\\
14. Yu.A.Simonov, {\it Yad.Fiz.} {\bf 58}, 113 (1995), {\it Lectures 
at the 35-th Internationale Universit\"atswochen f\"ur Kern- und 
Teilchenphysik, Schladming, March 2-9, 1996} (Springer-Verlag, in press).\\
15. Yu.A.Simonov, {\it Yad.Fiz.} {\bf 54}, 192 (1991).\\
16. V.N.Popov, {\it Functional Integrals in Quantum Field Theory
and Statistical Physics} (Reidel, 1983).\\
17. N.G.Van Kampen, {\it Stochastic Processes in Physics and
Chemistry} (North-Holland Physics Publishing, 1984).\\
18. Yu.M.Makeenko, {\it Phys.Lett.} {\bf B212}, 221 (1988), unpublished.\\
19. E.Witten, {\it Nucl.Phys.} {\bf B188}, 513 (1981); P.Salomonsen
and J.W.Van Holten, {\it Nucl.Phys.} {\bf B196}, 509 (1982); S.Cecotti
and L.Girardello, {\it Ann.Phys.} {\bf 145}, 81 (1983); C.Bender et
al., {\it Nucl.Phys.} {\bf B219}, 61 (1983).\\
20. E.Gozzi, {\it Phys.Rev.} {\bf D28}, 1922 (1983).\\
21. P.H.Damgaard and K.Tsokos, {\it Lett.Math.Phys.} {\bf 8}, 535 (1984).\\
22. J.Greensite and M.B.Halpern, {\it Nucl.Phys.} {\bf B211}, 343
(1983); M.B.Halpern, {\it Nucl.Phys.} {\bf B228}, 173 (1983).\\
23. A.A.Migdal, {\it Nucl.Phys.} {\bf B265 [FS15]}, 594 (1986).\\
24. A.A.Migdal, PUPT-1509 ({\it hep-th} /9411100).\\
25. G.Veneziano, {\it Nucl.Phys.} {\bf B117}, 519 (1976).\\
26. M.B.Halpern and Yu.M.Makeenko, {\it Phys.Lett.} {\bf B218}, 230
(1989).\\
27. M.Namiki et al., {\it Progr.Theor.Phys.} {\bf 69}, 1580 (1983).\\
\end{document}